\documentclass[aps,prl,twocolumn,groupedaddress,showpacs]{revtex4}
\usepackage{amsmath}
\usepackage{amsfonts}
\usepackage{amssymb}
\usepackage{graphicx}
\usepackage{epsfig}

\input{tcilatex}

\begin{document}

\title{Reentrant Behavior of the Spinodal Curve in a Nonequilibrium Ferromagnet}
\author{P.I. Hurtado}
\affiliation{Department of Physics, Boston University, Boston, MA 02215, USA \\
and Instituto Carlos I de F\'{\i}sica Te\'{o}rica y Computacional, Universidad de 
Granada, E-18071-Granada, Espa\~{n}a}
\author{J. Marro}
\author{P.L. Garrido}
\affiliation{Instituto Carlos I de F\'{\i}sica Te\'{o}rica y Computacional,\\
and Departamento de Electromagnetismo y F\'{\i}sica de la Materia,\\
Universidad de Granada, E-18071-Granada, Espa\~{n}a}
\date{\today}

\begin{abstract}
The metastable behavior of a kinetic Ising--like ferromagnetic model system
in which a generic type of microscopic disorder induces nonequilibrium
steady states is studied by computer simulation and a mean--field approach.
We pay attention, in particular, to the spinodal curve or intrinsic coercive
field that separates the metastable region from the unstable one. We find
that, under strong nonequilibrium conditions, this exhibits reentrant
behavior as a function of temperature. 
That is, metastability does not happen in this regime for both low and high 
temperatures, but instead emerges for intermediate temperature, as a consequence of
the non-linear interplay between thermal and nonequilibrium fluctuations.
We argue that this behavior, which is in contrast with equilibrium phenomenology and 
could occur in actual impure specimens, might be related to the presence of an
effective multiplicative noise in the system.
\end{abstract}

\pacs{05.20.Dd, 02.50.Ey, 05.70.Ln, 64.60.My}
\maketitle

\section{I. Introduction}

The concept of metastability \cite{Pnros,Gunton} is crucial to many branches
of science. Metastable states occur in liquids and glasses,\cite{Kob}
quark/gluon plasmas,\cite{quarks} globular proteins,\cite{proteinas}
cosmological phase transitions,\cite{cosmo} the \textquotedblleft standard
model\textquotedblright\ of particle physics,\cite{Isidori} climate models,%
\cite{clima} black holes and protoneutronic stars,\cite{estrellas} for
instance. Understanding metastability from a microscopic point of view is
therefore most interesting. It is also a difficult task, given that the
concept is a kinetic feature which is not described by the Gibbs ensemble
theory.\cite{Pnros} Consequently, a lot of activity still focuses on very
simple cases, particularly, kinetic Ising--like models. Some recent studies
along this line concern the details of nucleation during the relaxation
processes,\cite{Langer} some exact results in the limit of zero temperature,%
\cite{Schonmann} the checking of theoretical predictions by means of
computer simulation,\cite{simulaciones} and analysis of the effects of open
borders,\cite{Cirillo} quenched impurities,\cite{impurezas} and
demagnetizing fields.\cite{demag} 

These studies deal with systems in which the metastable state decays towards 
the \textit{equilibrium} stable state. In this case, some understanding 
can be achieved via nucleation theories in which Gibbs
thermodynamic (equilibrium) free energy plays a central role.
However, more general and intriguing is the case in which relaxation is
towards a \textit{nonequilibrium} steady state.\cite{Vacas,MD,nuestros,Pablotesis}
Nonequilibrium conditions appear ubiquitously in nature, and they characterize the 
evolution of most real systems.\cite{MD} Under such conditions, no free energy can
be defined in general,\cite{MD} and no coherent theoretical framework exists that
accounts for the observed far-from-equilibrium behavior. 
Some important questions regarding metastability concern 
the existence and properties of a nonequilibrium macroscopic potential capturing 
the essential physics of the metastable-stable transition under nonequilibrium 
conditions, and the limit of metastability when such conditions hold.


In this paper we therefore deal with aspects of metastability in a kinetic Ising--like
model with nonequilibrium steady states. Our interest is on the effects of
the nonequilibrium condition on the properties of the metastable state as
one varies the system parameters. In particular, we study the
magnetic--field strength for the onset of instability. This is the \textit{%
intrinsic coercive field }\cite{Intrinsic} which locates in magnets the 
\textit{spinodal curve} which is familiar from studies of density--conserved
systems.\cite{binder} The system behavior around this curve is the
consequence of a complex interplay between thermal and nonequilibrium
fluctuations. This results in a spinodal curve that depicts novel behavior as
compared to the equilibrium case. An interesting observation is that 
metastability occurs in the strong nonequilibrium regime at
intermediate temperatures but not in the low temperature limit, pointing out that,
in this regime, noise enhances metastability.

Recent studies on critical behavior of some nonequilibrium models have
predicted similar reentrance phenomena. That is, in a large class of model
systems, one observes that, under nonequilibrium fluctuations, a disordered
phase which characterizes the system at both low and high temperatures
becomes ordered at intermediate temperatures.\cite{MD,Thorpe,R1,R2,R2b,R3,R4} 
Further research is still needed,
however, before one may conclude on the relevance of such model behavior on
the reentrance phenomena reported, for instance, in nonequilibrium phase
transitions driven by competition between quantum and thermal fluctuations
in superconductors and vortex matter,\cite{RE1,RE2,RE3} and concerning
different liquid, glassy and amorphous phases in water and silica.\cite%
{REb1,REb2,REb3} Despite the similarities between our results and these studies, they
are different in essence: the latter concern reentrance in phase diagrams associated
to nonequilibrium phase transitions, while our work concern reentrance of the nonequilibrium
spinodal curve, which characterizes the limit of metastability.




The paper is organized as follows. Sections II and III describe,
respectively, the model and a dynamic mean--field approximation. Section IV
contains our main results on the static properties of nonequilibrium
metastability; in particular, we evaluate the intrinsic coercive field.
In Section V we measure this spinodal field in
computer simulations, and numerical results are compared in this section
with our mean--field theory. Section VI is devoted to conclusions.

\section{II. The Model}

Consider the two--dimensional Ising model on the square lattice of side $L,$ 
$\Lambda =\left\{ 1,\ldots ,L\right\} ^{2}\subset \mathbb{Z}^{2},$ with
periodic (toroidal) boundary conditions. There is a spin variable at each
lattice site with two possible states, $s_{i}=\pm 1,$ $i\in \Lambda ,$ and
spin--spin interactions and influence of an external magnetic field $h$ as
described by the function%
\begin{equation}
\mathcal{H}(\mathbf{s})=-\sum_{\langle i,j\rangle }s_{i}s_{j}-h\sum_{i\in
\Lambda }s_{i},  \label{hamilt}
\end{equation}%
where $\mathbf{s}\equiv \left\{ s_{i}\right\} $ and the first sum runs over
all nearest--neighbor pairs. Futhermore, the spin system evolves with time
via stochastic single--spin--flip dynamics as determined by the master
equation:%
\begin{equation}
\frac{\text{d}P(\mathbf{s};t)}{\text{d}t}=\sum_{i\in \Lambda }\left[ \omega (%
\mathbf{s}^{i}\rightarrow \mathbf{s})P(\mathbf{s}^{i};t)-\omega (\mathbf{s}%
\rightarrow \mathbf{s}^{i}) P(\mathbf{s};t)\right] .  \label{mastereq2}
\end{equation}%
Here, $P(\mathbf{s};t)$ is the probability of configuration $\mathbf{s}$ at
time $t,$ $\mathbf{s}^{i}$ stands for $\mathbf{s}$ after performing a flip
at $i,$ i.e., $s_{i}\rightarrow -s_{i},$ and $\omega (\mathbf{s}\rightarrow 
\mathbf{s}^{i})$ stands for the corresponding transition rate. 
In order to ensure nonequilibrium conditions, we introduce a weighted competition 
between two different temperatures (one ``infinite'' and the other finite).
This has been shown to be the simplest way of inducing nonequilibrium behavior in 
lattice models.\cite{MD} The rate is then chosen to be
\begin{equation}
\omega (\mathbf{s}\rightarrow \mathbf{s}^{i})=p+(1-p)\frac{\text{e}^{-\beta
\Delta \mathcal{H}(s_{i},n_{i})}}{1+\text{e}^{-\beta \Delta \mathcal{H}%
(s_{i},n_{i})}},  \label{rate}
\end{equation}%
where $\beta =1/T$ stands for the lattice (inverse) {\it temperature} ---so that
we are fixing the Boltzmann's constant to unity---, and $\Delta \mathcal{H}%
(s_{i},n_{i})\equiv \mathcal{H}(\mathbf{s}^{i})-\mathcal{H}(\mathbf{s}%
)=2s_{i}\left[ 2(n_{i}-2)+h\right] $, where $n_{i}\in \lbrack 0,4]$ is the
number of up nearest--neighbors of $s_{i}.$

\begin{figure}[tbp]
\centerline{
\psfig{file=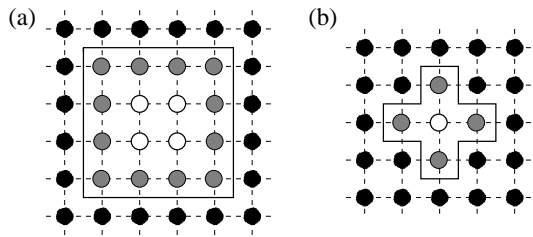,width=7cm}}
\caption[Different examples of spin domains.]{{\protect\small Two examples
of spin domains, each following from a different lattice partition $\mathbb{P%
}(\Lambda );$ see the main text. The spins that do not belong to the domain
are in black, \textit{surface} spins are gray, and the spins forming the
domain \textit{interior} are empty.}}
\label{dominio}
\end{figure}

The rate (\ref{rate}) describes spin flips under the action of
two competing heat baths. For $p=0$, the system goes asymptotically towards
the unique Gibbs, equilibrium state for temperature $T$ and energy $\mathcal{%
H}$. This has a critical point at $h=0$ and $T=T_{c}\left( p=0\right) =T_{%
\text{ons}}\approx 2.2691,$ the Onsager, equilibrium critical temperature.
For $0<p<1$, the conflict in (\ref{rate}) impedes canonical equilibrium,
and the system evolves asymptotically towards a nonequilibrium steady state
which may essentially differ from a Gibbs state.\cite{hamefectivo,MD} In this case,
no equilibrium thermodynamic global temperature can be defined. Now parameter $T$
can be thought as a source of thermal fluctuations, which compete with the 
non-thermal (nonequilibrium) noise induced by $p$. The system now
exhibits a critical point, at $h=0$ and $T=T_{c}\left( p\right)
<T_{c}\left( 0\right) ,$ which is apparently of the Ising universality class,%
\cite{Tamayo,FattahJJMA} but only as far as $p$ is small enough. In fact,
the nonequilibrium disorder which is implied by (\ref{rate}) washes out the
critical point for any $p>p_{c}\approx 0.17.$\cite{MD} One may think of the
dynamic random perturbation parameterized by $p$ as an extra source of
(nonequilibrium) disorder and randomness which is likely to occur also in
many actual systems in nature.\cite{Vacas,MD} 
A main question here is how the metastable states in the system depend on the competition 
between this non--thermal noise and the standard thermal fluctuations parameterized by $%
T. $

\section{III. A Mean--Field Approximation}

We first study a mean--field solution of (\ref{mastereq2}) in the pair
approximation.\cite{Dickmanpair,MD} This approach is a dynamic generalization
of Kikuchi's method known as \emph{Cluster Variation Method}.\cite{Kikuchi}
Consider a partition $\mathbb{P}$ of the
lattice such that resulting domains, $q_{j}\in \mathbb{P}(\Lambda ),$
satisfy $q_{j}\cap q_{j^{\prime }}=\varnothing $ if $j\neq j^{\prime },$ and 
$\bigcup_{j}q_{j}=\Lambda .$ We define the surface $\mathcal{S}_{j}$ of $%
q_{j}$ as the set of all its spins that have at least one nearest neighbor
outside the domain;\cite{interior} the rest is the interior, namely, $%
\mathcal{I}_{j}\equiv q_{j}-\mathcal{S}_{j},$ and it follows that $q_{j}=%
\mathcal{I}_{j}\cup \mathcal{S}_{j}.$ These definitions are illustrated in
Fig.\ref{dominio}. Next, consider a local observable $A(\mathbf{s}%
_{q_{j}};j) $ which exclusively depends on spins belonging to $q_{j}.$ One
readily has from (\ref{mastereq2}) for the average $\langle A(j)\rangle
_{t}\equiv \sum_{\mathbf{s}}A(\mathbf{s}_{q_{j}};j)P(\mathbf{s};t)$ that%
\begin{gather}
\frac{\text{d}\langle A(j)\rangle _{t}}{\text{d}t}=\sum_{\mathbf{s}%
_{q_{j}}}\sum_{i\in \mathcal{I}_{j}}\Delta A(\mathbf{s}_{q_{j}};j;i)\omega (%
\mathbf{s}_{q_{j}}\rightarrow \mathbf{s}_{q_{j}}^{i})Q(\mathbf{s}_{q_{j}};t)
\notag \\
+\sum_{\mathbf{s}}\sum_{i\in \mathcal{S}_{j}}\Delta A(\mathbf{s}%
_{q_{j}};j;i)\omega (\mathbf{s}\rightarrow \mathbf{s}^{i})P(\mathbf{s};t),
\label{exacta}
\end{gather}%
where $\mathbf{s}_{q_{j}}$ is the configuration of the domain spins, $\Delta
A(\mathbf{s}_{q_{j}};j;i)=A(\mathbf{s}_{q_{j}}^{i};j)-A(\mathbf{s}%
_{q_{j}};j),$ and $Q(\mathbf{s}_{q_{j}};t)\equiv \sum_{\mathbf{s}-\mathbf{s}%
_{q_{j}}}P(\mathbf{s};t)$ is the probability of having the configuration $%
\mathbf{s}_{q_{j}}$ at time $t$. The notation $\omega (\mathbf{s}%
_{q_{j}}\rightarrow \mathbf{s}_{q_{j}}^{i})$ in eq. (\ref{exacta}) stresses
the fact that flipping a spin in the interior only depends on the spins
belonging to the domain.

Let us assume that the system is \textit{spatially homogeneous}, namely,
that $\langle A(j)\rangle \equiv \langle A\rangle ,$ $q_{j}\equiv q,$ $%
\mathcal{I}_{j}\equiv \mathcal{I},$ and $\mathcal{S}_{j}\equiv \mathcal{S}$
for any $j.$ Equivalently, the partition $\mathbb{P}(\Lambda )$ is regular,
so that all domains are topologically identical. One notices that the two
r.h.s. terms in eq.(\ref{exacta}) concern the domain interior and surface,
respectively; the latter couples the domain dynamics to its surroundings.
Our second approximation consists in neglecting this surface term, i.e., any
correlation during time evolution which extends outside the domain. Under
these two approximations, homogeneity and kinetic isolation, equation (\ref%
{exacta}) reduces to%
\begin{equation}
\frac{\text{d}\langle A\rangle _{t}}{\text{d}t}=\sum_{\mathbf{s}%
_{q}}\sum_{i\in \mathcal{I}}\Delta A(\mathbf{s}_{q};i)\omega (\mathbf{s}%
_{q}\rightarrow \mathbf{s}_{q}^{i})Q(\mathbf{s}_{q};t).  \label{pairapprox}
\end{equation}

Next, one needs to estimate $Q(\mathbf{s}_{q};t)$ in terms of $n-$body
correlation functions. Assuming that only $\langle s\rangle $ and $\langle
s_{i}s_{j}\rangle $, with $i$ and $j$ nearest--neighbor sites inside the
domain, matter, $Q(\mathbf{s}_{q};t)$ may be written as a function of the
spin density $\rho (s)$ and the density $\rho (s,s^{\prime })$ of
nearest-neighbors pairs only. Furthermore, as only nearest--neighbors
correlations are taken into account, we consider a domain with only one spin
in its interior, which has $4$ surface spins; see Fig.\ref{dominio}b. 
With this choice, our mean-field theory turns out to be a nonequilibrium analog 
of the equilibrium Bethe-Peierls approximation.
It follows that the probability of finding the central spin in state $s$
surrounded by $n$ up nearest--neighbor spins is%
\begin{equation}
Q(\mathbf{s}_{q};t)\equiv Q(s,n)={\binom{4}{n}}\frac{\rho (+,s)^{n}\rho
(-,s)^{4-n}}{\rho (s)^{3}}.  \label{binomial2}
\end{equation}%
Therefore, using the relations $\rho (+,-)=\rho (-,+)=\rho (+)-\rho (+,+)$
and $\rho (-,-)=1+\rho (+,+)-2\rho (+)$, and writing $x\equiv \rho (+)$ and $%
z\equiv \rho (+,+)$, eq. (\ref{pairapprox}) reads 
\begin{gather}
\frac{\text{d}\langle A\rangle _{t}}{\text{d}t}=\sum_{n=0}^{4}{\binom{4}{n}}%
\left[ \Delta A(+,n)\frac{z^{n}(x-z)^{4-n}}{x^{3}}\omega (+,n)\right.  \notag
\\
-\Delta A(-,n)\left. \frac{(x-z)^{n}(1+z-2x)^{4-n}}{(1-x)^{3}}\omega (-,n)%
\right] ,  \label{pairapprox2}
\end{gather}%
where $\omega (s,n)\equiv \omega (\mathbf{s}_{q}\rightarrow \mathbf{s}%
_{q}^{i})$. This is for local isotropic observables for which the dependence
on $\mathbf{s}_{q}$ is through the pair $(s,n)$ only, $A(\mathbf{s}%
_{q};t)\equiv A(s,n).$

One may apply (\ref{pairapprox2}) to the observables $A_{1}(s,n)=\frac{1}{2}%
\left( 1+s\right) $ and $A_{2}(s,n)=\frac{1}{8}n\left( 1+s\right) $ whose
averages are $x$ and $z,$ respectively. Then, $\Delta A_{1}(s,n)=-s,$ $%
\Delta A_{2}(s,n)=-%
{\frac14}%
sn,$ and%
\begin{eqnarray}
\frac{\text{d}x}{\text{d}t} &=&F_{1}(x,z)\equiv \sum_{n=0}^{4}G\left(
x,z;n\right)  \label{xzpair1} \\
\frac{\text{d}z}{\text{d}t} &=&F_{2}(x,z)\equiv \sum_{n=0}^{4}\frac{n}{4}%
G\left( x,z;n\right) ,  \label{xzpair2}
\end{eqnarray}

where%
\begin{eqnarray*}
G\left( x,z;n\right) &\equiv &{\binom{4}{n}}\left[ \frac{%
(x-z)^{n}(1+z-2x)^{4-n}}{(1-x)^{3}}\omega (-,n)\right. \\
&&-\left. \frac{z^{n}(x-z)^{4-n}}{x^{3}}\omega (+,n)\right] .
\end{eqnarray*}%
Together with (\ref{rate}) and (\ref{pairapprox2}), these equations provide $%
x(t)$ and $z(t)$ as well as any other local isotropic magnitude.
\begin{figure}[tbp]
\centerline{
\psfig{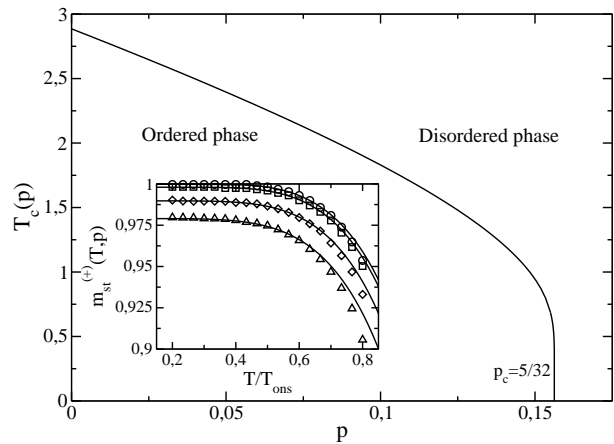}}
\caption[Phase diagram of the nonequilibrium magnetic system.]{%
{\protect\small Variation with $p$ of the critical {\it temperature} for the
nonequilibrium ferromagnetic system in the first--order mean--field
approximation. The solid line in the inset stands for the locally--stable
steady magnetization as a function of {\it temperature} (in units of the
equilibrium critical value) for $h=0$ and, from top to bottom, $p=0$, $0.001$%
, $0.005$ and $0.01$. The symbols in the inset are Monte Carlo results for a 
$53\times 53$ lattice. }}
\label{phasedia}
\end{figure}

\section{IV. Static Properties}

Our main interest here is on the steady solutions%
\begin{equation}
F_{1}(x_{st},z_{st})=0,\qquad F_{2}(x_{st},z_{st})=0,  \label{f1f2}
\end{equation}%
and on their stability.\cite{note1} Both stable and metastable states are
locally stable under small perturbations, which requires the (necessary and
sufficient) condition:\cite{Hurwitz}%
\begin{gather}
\left( \frac{\partial F_{1}}{\partial x}\right) _{st}+\left( \frac{\partial
F_{2}}{\partial z}\right) _{st}<0,\qquad \text{and}  \notag \\
\left( \frac{\partial F_{1}}{\partial x}\right) _{st}\left( \frac{\partial
F_{2}}{\partial z}\right) _{st}-\left( \frac{\partial F_{1}}{\partial z}%
\right) _{st}\left( \frac{\partial F_{2}}{\partial x}\right) _{st}>0.
\label{Hurwitz}
\end{gather}%
The condition%
\begin{equation}
\left[ \frac{\partial F_{1}(x,z)}{\partial x}\right] _{st}=0,
\label{condcrit}
\end{equation}%
on the other hand, characterizes incipient or marginal stability, i.e., the
presence of a critical point $(x_{st}^{c},z_{st}^{c})$ for $h=0,$ $%
x_{st}^{c}=\frac{1}{2}$ and $z_{st}^{c}=\frac{1}{3},$ and it follows that%
\begin{equation}
T_{c}(p)=-4\left[ \ln \left( \frac{3}{4}\sqrt{\frac{1-4p}{1-p}}-\frac{1}{2}%
\right) \right] ^{-1}.  \label{tempBethe}
\end{equation}%
This is the critical {\it temperature} for the nonequilibrium model in the present
first--order mean--field approximation;\cite{MD} see Fig.\ref{phasedia}. It
is noticeable the existence of a critical value of $p$ such that $%
T_{c}(p_{c})=0,$ which gives $p_{c}=5/32=0.15625$ (to be compared with the 
\textit{exact} value $p_{c}\simeq 0.17)$. 

The stationary state $(x_{st},z_{st})$ may be obtained numerically from the
non--linear differential equations (\ref{f1f2}) subject to the local
stability condition (\ref{Hurwitz}). For $h=0,$ the up--down symmetry leads
to pairs of locally--stable steady solutions, namely $(x_{st},z_{st})$ and $%
(1-x_{st},1+z_{st}-2x_{st})$. The result is illustrated in the inset of Fig. %
\ref{phasedia}; this also shows a comparison with Monte Carlo results which
confirms the expected agreement at low and intermediate temperatures for any 
$p$. The fact that increasing $p$ at fixed $T$ decreases the magnetization
implies that the nonequilibrium perturbation tends to increase
disorder. For small enough fields, the situation closely resembles the case $%
h=0;$ the up--down symmetry is now broken, however, and locally--stable
steady states with magnetization oriented opposite to the applied field are
metastable. 

\begin{figure}[t]
\centerline{
\psfig{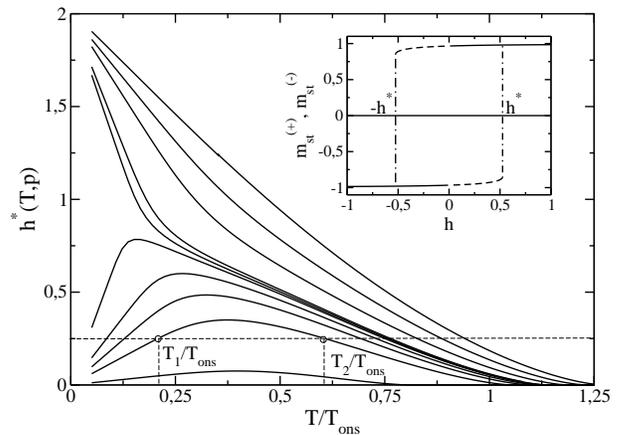}}
\caption[Intrinsic coercive field as obtained from mean field approximation.]%
{{\protect\small $h^{\ast }(T,p)$, as a
function of $T$ for, from top to bottom, $p=0$, $0.01$, $0.02$, $%
0.03 $, $0.031$, $0.032$, $0.035$, $0.04$, $0.05$ and $0.1$. The qualitative
change of behavior in the low temperature region occurs for $p\in
(0.031,0.032)$. 
Inset: The two locally--stable steady magnetization branches as a function of 
$h$ for $T=0.7T_{c}(0)$ and $p=0.005$. The solid (dashed) line represents stable 
(metastable) states. The dot-dashed line signals
the discontinuous transition, at $h=h^{\ast}(T,p)$, where metastable states 
disappear.}}
\label{campocritico}
\end{figure}

The locally-stable steady magnetization exhibits two 
branches as a function of $h$; see inset in Fig. \ref{campocritico}.
This hysteresis loop reveals that metastability does not occur for
any $\left\vert h\right\vert >h^{\ast }(T,p)\geq 0,$ where $h^{\ast }(T,p)$
is the \textit{intrinsic coercive field}.\cite{icf} 
Let $z=z(x)$ the solution of eq. (\ref{xzpair2}), and write eq. 
(\ref{xzpair1}) as 
\begin{equation}
\frac{\text{d}x}{\text{d}t}=-\frac{\delta V(x)}{\delta x}.  \label{potencial}
\end{equation}%
This defines $V(x),$ a (nonequilibrium) bimodal potential that controls the system 
time evolution. An increase in the field tends to attenuate the local minimum 
in $V(x)$ associated with the metastable state. This minimum exists only for 
$|h|<h^{\ast }(T,p)$; the set of eqs.(\ref{f1f2}) has only one solution, with
magnetization sign equal to that of the applied field, for $|h|>h^{\ast
}(T,p)$. 

In order to evaluate $h^{\ast }(T,p),$ one may study how the metastable
state responds to small perturbations of the applied field. If
$\left( x_{st}^{h_{0}},z_{st}^{h_{0}}\right) $ is a locally--stable
stationary state for $T$, $p$ and $h_{0}$, with magnetization opposed to $h$, 
and we perturb $h=h_{0}+\delta h$, the new locally--stable stationary solution
is modified: $x_{st}^{h}=x_{st}^{h_{0}}+\epsilon _{x}$ and $%
z_{st}^{h}=z_{st}^{h_{0}}+\epsilon _{z}$. One obtains at first order that 
\begin{equation}
\epsilon _{x}=\left[ \frac{\frac{\partial F_{2}}{\partial h}\frac{\partial
F_{1}}{\partial z}-\frac{\partial F_{1}}{\partial h}\frac{\partial F_{2}}{%
\partial z}}{\frac{\partial F_{1}}{\partial x}\frac{\partial F_{2}}{\partial
z}-\frac{\partial F_{2}}{\partial x}\frac{\partial F_{1}}{\partial z}}\right]
_{x_{st}^{h_{0}},z_{st}^{h_{0}},h_{0},T,p}\delta h,  \label{respuesta}
\end{equation}%
This (linear) response diverges for%
\begin{equation}
\left[ \frac{\partial F_{1}}{\partial x}\frac{\partial F_{2}}{\partial z}-%
\frac{\partial F_{2}}{\partial x}\frac{\partial F_{1}}{\partial z}\right]
_{x_{st}^{h_{0}},z_{st}^{h_{0}},h_{0},T,p}=0,  \label{condh}
\end{equation}%
which corresponds to a discontinuity in the metastable magnetization as a
function of $h$. For fixed $T$ and $p$, the field for which (\ref{condh})
holds is $h^{\ast }(T,p)$. Fig. \ref{campocritico} shows
the mean-field result for $h^{\ast }(T,p)$. In particular, for $p=0$ (the equilibrium 
case) we recover the standard equilibrium mean-field spinodal curve: converging to $2$
as $T\rightarrow 0$, linearly decreasing with temperature for small $T$, and vanishing 
as $(T_c-T)^{3/2}$ at the mean-field equilibrium critical point. A main result derived 
from Fig. \ref{campocritico} is the existence of two 
different low temperature limits for $h^{\ast }(T,p)$. For small enough values
of $p,$ namely, $p\in \lbrack 0,0.031],$ which includes the equilibrium
case, $p=0,$ the field $h^{\ast }(T,p)$ monotonously grows and 
extrapolates to $2$ as $T\rightarrow0.$ 
For larger $p$, namely, $p\in \lbrack 0.032,\frac{5}{32}),$ however, $%
h^{\ast }(T,p)\rightarrow 0$ as $T\rightarrow 0$, exhibiting a maximum at 
intermediate $T$. The value $p=\pi_{c}\approx 0.0315$ separates the two types of 
asymptotic behavior.

When we cool the system in the regime $p<\pi _{c}$, the field $h^{\ast }(T,p)$ increases,
so in this case we require a stronger field to \textit{destroy} the metastable state.
This may be understood on simple grounds. The tendency of
spins to line up in the direction of the field competes with the tendency to
maintain order implied by their mutual interactions. A metastable state
lasts for a long time because the latter prevails over the action of the
field. Both $T$ and $p$ induce disorder; therefore, lowering $T$ increases
order, so that a stronger field is needed to destroy the metastable state,
which is in fact observed for $p<\pi _{c}$. On the other hand, as $p$ is
increased, the disorder increases, and $h^{\ast }(T,p)$ needs to decrease
for a fixed $T,$ according to our observations. 
\begin{table}[t]
\centerline{
\begin{tabular}{|c||c|c|c|}
\hline \hline
Class & Central spin & Number of up neighbors & $\Delta {\cal H}$ \\
\hline \hline
1 & +1 & 4 & 8J+2h \\
\hline
2 & +1 & 3 & 4J+2h \\
\hline
3 & +1 & 2 & 2h \\
\hline
4 & +1 & 1 & -4J+2h \\
\hline
5 & +1 & 0 & -8J+2h \\
\hline \hline
6 & -1 & 4 & -8J-2h \\
\hline
7 & -1 & 3 & -4J-2h \\
\hline
8 & -1 & 2 & -2h \\
\hline
9 & -1 & 1 & 4J-2h \\
\hline
10 & -1 & 0 & 8J-2h \\
\hline \hline
\end{tabular}
}
\caption[Spin classes for the two-dimensional Ising model.]{%
{\protect\small Spin classes for the two-dimensional Ising model
with periodic boundary conditions. The last column shows the energy cost of
flipping the central spin.}}
\label{table}
\end{table}

The picture for $p>\pi _{c}$ is more intriguing. Consider the case $%
|h|=0.25$ and $p=0.05>\pi _{c}$. As illustrated in Fig.\ref{campocritico},
one may define two {\it temperatures}, $T_{1}<T_{2},$ such that metastable states
only occur for $T\in (T_{1},T_{2}).$ The fact that $h^{\ast }(T,p)$
extrapolates to zero in the low temperature limit for $p=0.05>\pi _{c}$
indicates that such amount of nonequilibrium noise is able to destroy on its own 
any metastability. Following the above reasoning, increasing $T$ adds disorder,
so that no metastability should, in principle, show up in this case.
However, there is a regime of intermediate temperatures, $T\in (T_{1},T_{2})$%
, for which metastability occurs. This noise-enhanced metastability 
is a consequence of the complex
interplay between the standard thermal fluctuations and nonequilibrium 
noise: although both noises add independently disorder, their
combination determines the existence of regions in the parameter space $%
(T,p) $ in which no metastable states occur at low $T$ but only at intermediate
{\it temperatures}. This reentrance phenomenon is reminiscent of the one
observed in the annealed Ising model \cite{Thorpe} and other closely related
systems \cite{MD,R1,R2,R2b,R3,R4} where multiplicative noise seems to play an
essential role.\cite{R2b}

\section{V. Monte Carlo Simulations: Growth and Shrinkage of the Stable Phase}

In this section, we check further our theoretical predictions against
computer simulation data. With this aim, we need a simple criterion to
conclude that the model system exhibits metastable states. Let us first characterize
all the possible local configurations in terms of the spin state, $s=\pm 1,$
and the number of its up nearest neighbors, $n\in \lbrack 0,4]$. For
periodic boundary conditions, there are $10$ different \textit{spin classes,} as 
shown in Table \ref{table}. The cost $\Delta \mathcal{H}(s,n)$ of flipping any spin
within a class is the same. That is, the rate (\ref{rate}) only depends
on $s$ and $n,$ which define the class. If $n_{k}({\bf s})$ is the number of spins
in class $k$ when the system is is configuration ${\bf s}$, and noticing that 
classes $k\in \lbrack 1,5]$ are characterized by a central up spin,
we may write the number of up spins which flip per unit time in the state ${\bf s}$ as 
\begin{equation}
G({\bf s})=\sum_{k=1}^{5}n_{k}({\bf s})\omega _{k}.  \label{growing}
\end{equation}%
As far as $h<0$, this is the \textit{growth rate} of the stable phase in state 
${\bf s}$. The \textit{shrinkage rate} of the stable phase
follows similarly as,\cite{projective} 
\begin{equation}
S({\bf s})=\sum_{k=6}^{10}n_{k}({\bf s})\omega _{k}.  \label{shrinking}
\end{equation}%
Given a phase-space point ${\bf s}$, the rates $G({\bf s})$ and $S({\bf s})$ yield the 
{\it local} tendency of the system to evolve toward the stable or metastable phases, 
respectively. 

For $h<0$, a state with all spins up, ${\bf s_1} \equiv \{s_i=+1, i=1,\ldots,N\equiv L^2 \}$, 
will relax after some time toward the stable steady state, which corresponds in this case to a 
configuration with negative magnetization, $m<0$. For a given experiment $j$ of a total of 
$N_{exp}$ runs, this relaxation will proceed through certain path in phase space, which we note as 
$\sigma_j \equiv \{{\bf s_1}, {\bf s_2^{(j)}}, \ldots, {\bf s_{\Gamma(j)}^{(j)}} \}$.
Here ${\bf s_l^{(j)}}$ is the $l$-th configuration, starting from ${\bf s_1}$, of a total number of 
$\Gamma(j)$ configurations which make up the path in experiment $j$. At any stage ${\bf s_l}^{(j)}$ of
this path, the difference $G({\bf s_l^{(j)}})-S({\bf s_l^{(j)}})$ defines the net tendency of the system
to evolve toward the final steady stable state. A metastable state is characterized by the presence of 
{\it free energy} barriers hampering the relaxation toward the truly 
stable state. In this case, relaxation is an activated process controled by large fluctuations. On
the other hand, an unstable state decays without any hindrance. Therefore, we may divide relaxation
paths in two different types. On one hand, {\it metastable-like paths}, in which at least one 
configuration ${\bf s_k^{(j)}}\in \sigma_j$ exists, excluding the final stable one, such that 
$G({\bf s_k^{(j)}})-S({\bf s_k^{(j)}})<0$, and on the other hand, {\it unstable-like paths}, 
such that $G({\bf s_k^{(j)}})-S({\bf s_k^{(j)}})>0$ $\forall {\bf s_k^{(j)}}\in \sigma_j$, 
excluding again the final stable state.

\begin{figure}[t]
\centerline{
\psfig{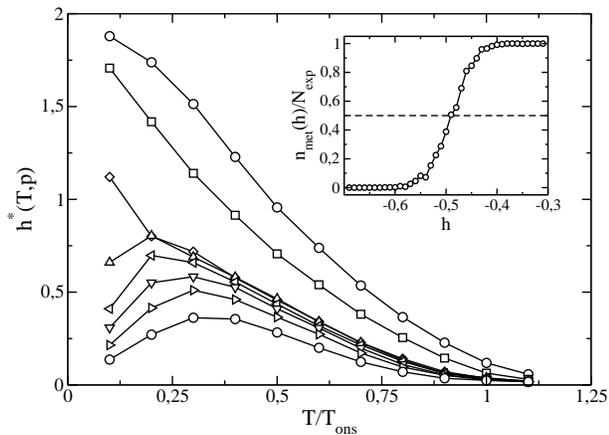}}
\caption[Intrinsic coercive 
field as obtained from Monte Carlo simulations.]{{\protect\small %
Monte Carlo results
for $h^{\ast}(T,p)$ as a function of $T$ for
$L=53$ and, from top to bottom, $p=0$, $0.01$, $0.03$, $0.0305$, $0.0320$,
$0.0350$, $0.04$ and $0.05$. Notice the change of asymptotic behavior in the
low temperature limit for $p\in(0.03,0.0305)$. Inset: The probability of the 
metastable state, as defined in the main text, as a function 
of $h<0$ for $L=53$, $T=0.7T_{c}(0)$ and $p=0.$ Data here 
correspond to an average over $500$ independent demagnetization experiments for each 
value of $h$. Error bars are smaller than the symbol sizes.}}
\label{hcritMC2}
\end{figure}
For fixed $T$, $p$ and $h<0$, given the stochasticity of the dynamics, 
one needs to be concerned with the 
\textit{probability} of occurrence of metastability, defined as
$n_{met}(T,p,h)/N_{exp}$, where $n_{met}(T,p,h)$ is the number of experiments
out of the total $N_{exp}$ in which the relaxation path in phase space belongs to 
the class of {\it metastable-like paths}.
This is shown in the inset of Fig.\ref{hcritMC2}. The
intrinsic coercive field, $h^{\ast }(T,p)$, is defined in this scheme as the field for 
which $n_{met}(T,p,h^{\ast })/N_{exp}=0.5;$ this is shown in Fig. \ref{hcritMC2}
for a system with $L=53$.

A detailed comparison of these numerical results with the theory in Fig.\ref%
{campocritico} depicts semi--quantitative agreement, namely, the agreement
is excellent except ---as one should have expected--- near the critical
temperature. In particular, the numerical critical value $\pi _{c}$ for $p$
is $\pi _{c}^{MC}\approx 0.03025$, rather close to the theoretical
prediction $\pi _{c}^{pair}\approx 0.0315$. This nicely confirm that the
addition of sufficient thermal noise in the presence of a large enough
nonequilibrium perturbation, $p>\pi _{c},$ tends to restore metastability.
We have also looked for finite-size corrections to the measured spinodal field by
simulating larger systems, finding that these corrections are very small, and can be
neglected for all practical purposes.

Finally, let us remark that the stable phase growth and shrinkage rates have been
introduced in literature as projected on the slow observable characterizing the relaxation 
process, namely the system magnetization $m$.\cite{projective} In this case, the rate $G(m)$ 
($S(m)$) yields the average number of up (down) spins which flip per unit time when magnetization 
is $m$. One may then define
\begin{eqnarray}
G(m)=\sum_{\{{\bf s} | m\}} {\cal P}({\bf s}) G(s), \quad
S(m)=\sum_{\{{\bf s} | m\}} {\cal P}({\bf s}) S(s),
\label{ratesprojected}
\end{eqnarray}
where $\{{\bf s} | m\}$ are all system configurations with fixed magnetization $m$, and 
${\cal P}({\bf s})$ is the probability of observing a configuration ${\bf s}$ during
the relaxation from the initial state, ${\bf s_1}$, toward the stable one.  
Steady states are usually determined by the condition $G(m)=S(m)$. However, the lack of
intersection between the curves $G(m)$ and $S(m)$ for $h<0$ in the $m>0$ interval does not
contain information about the limit between metastable and unstable states.\cite{referee} 
Instead, the magnetic field for which such lack of intersection first develops defines the 
so-called Dynamic Spinodal field, $h_{DSP}(T,p)$, which divides the metastable region of parameter
space $(T,p,h)$ for finite systems in two subregions characterized by different relaxation
morphologies.\cite{DSP} In particular, for $|h|<h_{DSP}(T,p)$ the metastable state relax through 
the nucleation of a single droplet of the stable phase, while for $h_{DSP}(T,p)<|h|<h^*(T,p)$
the relaxation proceeds via the nucleation of multiple stable-phase droplets.

\section{VI. Conclusion}

This paper deals with some of the static properties of metastable states in
a nonequilibrium Ising--like ferromagnetic model system, as obtained from
first--order mean--field theory and computer (Monte Carlo) simulations. We
studied, in particular, the spinodal or intrinsic coercive field, $h^{\ast
}(T,p)$, defined as the magnetic field strength for which the metastable state
becomes unstable. Our theoretical approximation predicts reentrace phenomena
as a function of $T$ in the strong nonequilibrium regime, $p>\pi _{c}\approx
0.0315$, where $p$ controls the dynamic, non--equilibrium perturbation. More
specifically, metastability is not observed at low temperatures for $p>\pi
_{c}$, but it occurs as one increases the {\it temperature}. This
noise--enhanced metastability reveals the existence of a complex interplay
between the thermal and nonequilibrium noises. That is, adding the
two effects ---which, independently, tend to increase disorder--- not always
results in decreasing the system ordering. The above is fully confirmed in
computer simulations, in which $h^{\ast }$ may accurately be measured from
the stable phase growth and shrinkage rates.

The physical origin of the observed noise--driven metastability is
intriguing. A clue to understand this behavior is to notice that certain
systems under the action of a multiplicative noise exhibit a similar
reentrant behavior, namely, disorder is dominant at the low and high
temperature regimes while well--defined order sets in at intermediate
temperatures. That the competition between thermal and nonequilibrium
fluctuations in (\ref{rate}) may induce an effective multiplicative noise can be understood
on simple grounds. The effect of the nonequilibrium perturbation in our model may be
described by means of an effective temperature $T_{eff}$, which is
{\it inhomogeneous} throughout the system for any $p>0$.\cite{Pablotesis} 
In fact, for any $\Delta{\cal H} \neq 0$ we may write (\ref{rate}) as an 
{\it equilibrium} Glauber rate with effective parameters, 
$\omega \equiv \textrm{exp}(-\beta_{eff}\Delta {\cal H})/[1+\textrm{exp}(-\beta_{eff}\Delta {\cal H})]$,
and this defines an effective temperature\cite{Metropolis}
\begin{equation}
T_{eff}(s,n) \equiv \frac{\Delta {\cal H}(s,n)}{\ln \big[\frac{1}{p+(1-p)
\frac{\textrm{e}^{-\beta \Delta{\cal H}(s,n)}}{1+\textrm{e}^{-\beta \Delta{\cal H}(s,n)}}} -1 \big]}
\label{teff}
\end{equation}
As a matter of fact, the temperature a spin effectively feels depends on the local 
order around it, i.e., on the number of nearest neighbors pointing in the same direction;
$T_{eff}$ is an increasing function of local order and,
consequently, the amplitude of the fluctuations depends on the local order
parameter. This is a main feature of Langevin--type models with a
multiplicative noise.\cite{R2b} 
Developing further this possibility to treat the limit of 
metastability is an open question.
This work, which seems most interesting, is outside the scope of present paper.

\section*{Acknowledgments}

We acknowledge very useful discussions with Miguel \'{A}ngel Mu\~{n}oz, and
financial support from MCYT--FEDER, project BFM2001-2841. P.I.H. also acknowledges
support from MECD.


\begin{thebibliography}{99}
\bibitem{Pnros} O. Penrose and J.L. Lebowitz, 
in \textit{Fluctuation Phenomena} (2nd edition), edited by E. Montroll and
J.L. Lebowitz, North$-$Holland, Amsterdam 1987.

\bibitem{Gunton} J.D. Gunton and M. Droz, \textit{Introduction to the Theory
of Metastable and Unstable States}, Springer, Berlin 1983.

\bibitem{Kob} W. Kob, 
\textit{J. Phys.:Condens. Mat.} \textbf{11}, R85 (1999)

\bibitem{quarks} E.E. Zabrodin, L.V. Bravina, H. St\"{o}cker, and W.
Greiner, 
\textit{Phys. Rev. C} \textbf{59}, 894 (1999)

\bibitem{proteinas} R.P. Sear, 
\textit{J. Chem. Phys.} \textbf{111}, 4800 (1999)

\bibitem{cosmo} A. Strumia and N. Tetradis, 
\textit{J. High Energy Phys.} \textbf{11}, 023 (1999)

\bibitem{Isidori} G. Isidori, G. Ridolfi, and A. Strumia, 
\textit{Nuc. Phys. B} \textbf{609}, 387 (2001)

\bibitem{clima} N. Berglund and B. Gentz, 
\textit{Stoch. and Dynam. \textbf{2}}, 327 (2002)

\bibitem{estrellas} S. Banik and D. Bandyopadhyay, 
\textit{Phys. Rev. C} \textbf{63}, 35802 (2001)

\bibitem{Langer} J.S. Langer, 
in \textit{Solids Far from Equilibrium}, edited by C. Godr{\`{e}}che,
Cambridge University Press, Cambridge 1992.

\bibitem{Schonmann} R.H. Schonmann, 
\textit{Commun. Math. Phys.} \textbf{147}, 231 (1992)

\bibitem{simulaciones} R.A. Ramos, P.A. Rikvold and M.A. Novotny, 
\textit{Phys. Rev. B} \textbf{59}, 9053 (1999)

\bibitem{Cirillo} E.N.M. Cirillo and J.L. Lebowitz, 
\textit{J. Stat. Phys.} \textbf{90}, 211 (1998)

\bibitem{impurezas} M. Kolesik \textit{et al.}, 
J. Appl. Phys. \textbf{81}, 5600 (1997)

\bibitem{demag} H.L. Richards, M.A. Novotny, and P.A. Rikvold, 
Phys. Rev. B \textbf{54}, 4113 (1996)

\bibitem{Vacas} J. Marro and J.A. Vacas, 
\textit{Phys. Rev. B} \textbf{56}, 8863 (1997)

\bibitem{MD} J. Marro and R. Dickman, \textit{Nonequilibrium Phase
Transitions in Lattice Models}, Cambridge University Press, Cambridge 1999.

\bibitem{nuestros} P.I. Hurtado, J. Marro and P.L. Garrido, \textit{AIP
Conference Proceedings} \textbf{661}, 147 (2003)

\bibitem{Pablotesis} P.I. Hurtado, J. Marro, and P.L. Garrido, to be
published.


\bibitem{Intrinsic} C. Rudowicz and H.W.F. Sung, \textit{Am. J. Phys., }to
appear.

\bibitem{binder} K. Binder, M.H. Kalos, J.L. Lebowitz, and J. Marro, \textit{%
Advances in Colloid and Interface Science} \textbf{10}, 173 (1979)

\bibitem{Thorpe} M.F. Thorpe and D. Beeman, \textit{Phys. Rev. B} \textbf{14}%
, 188 (1976)

\bibitem{R1} C. van den Broeck, J.M.R. Parrondo, and R. Toral, \textit{Phys.
Rev. Lett.} \textbf{73}, 3395 (1994); \textit{ibid}, \textit{Phys. Rev. E }%
\textbf{55},4084 (1997)

\bibitem{R2} J.J. Torres, P.L. Garrido, and J. Marro, \emph{Phys. Rev. B }%
\textbf{58}, 11488 (1998)

\bibitem{R2b} W. Genovese, M.A. Mu\~{n}oz and P.L. Garrido, 
\textit{Phys. Rev. E} \textbf{58}, 6828 (1998)

\bibitem{R3} S. Mangioni, R. Deza, R. Toral, and H. Wio, \textit{Phys.Rev. E 
}\textbf{61}, 223 (2000)

\bibitem{R4} M. Iba\~{n}es, J. Garcia-Ojalvo, R. Toral, and J.M. Sancho 
\textit{Phys. Rev. Lett.} \textbf{87}, 20601 (2001)

\bibitem{RE1} V.Yu. Butko, P.W. Adams, and E. I. Meletis, \emph{Phys. Rev.
Lett.}\textbf{\ 83}, 3725 (1999)%

\bibitem{RE2} K. Maki and S. Haas, \emph{Phys. Lett. A} \textbf{272}, 271
(2000)

\bibitem{RE3} R. B\"{u}gel, A. Fai\ss t, H.v. L\"{o}hneysen, J. Wosnitza,
and U. Schotte, \emph{Phys. Rev. B }\textbf{65}, 052402 (2001)%

\bibitem{REb1} L. Berthier, L.F. Cugliandolo and J.L. Iguain, \emph{Phys.
Rev. E} \textbf{63}, 051302 (2001)%

\bibitem{REb2} G. Franzese, G. Malescio, A. Skibinsky, S.V. Buldyrev, and H.
E. Stanley, \emph{Nature} \textbf{409}, 692 (2001)%

\bibitem{REb3} E.A. Jagla, \emph{Phys. Rev. E }\textbf{63}, 061509 (2001)%

\bibitem{hamefectivo} P.L. Garrido and J. Marro, \textit{Phys. Rev. Lett.} 
\textbf{62}, 1929 (1989); P.L Garrido and M.A. Mu\~{n}oz, 
\textit{Phys. Rev. E} \textbf{48}, R4153 (1993)

\bibitem{Tamayo} P. Tamayo, F.J. Alexander and R. Gupta, 
\textit{Int. J. Mod. Phys. C} \textbf{7}, 389 (1996)

\bibitem{FattahJJMA} A. Achahbar, J.J. Alonso and M.A. Mu\~{n}oz, 
\textit{Phys. Rev. E} \textbf{54}, 4838 (1996)

\bibitem{Dickmanpair} R. Dickman, \textit{Phys. Lett. A} \textbf{122}, 463
(1987)

\bibitem{Kikuchi} R. Kikuchi, \textit{Phys. Rev.} \textbf{81}, 988 (1951).

\bibitem{interior} More generally, $\mathcal{S}_{j}$ consists of all the
spins whose flipping probability depends on spins outside the domain.
However, in our case, in which the flipping probability depends on the value
of the spin and its four nearest neighbors, this reduces to the definition
in the main text.

\bibitem{note1} Our theoretical squeme cannot be used to study the evolution
from the metastable towards the stable state; this is a highly inhomogeneous
process triggered by fluctuations which escapes the present approach.

\bibitem{Hurwitz} L. Elsgotz, \textit{Ecuaciones Diferenciales y C\'{a}lculo
Variacional}, editorial MIR (1992)

\bibitem{icf} The intrinsic coercive field is defined as the magnetic field
for which the magnetization becomes zero in the hysteresis loop.\cite%
{Intrinsic} The loop in our case goes discontinuously from positive to
negative magnetization, and viceversa, crossing (discontinuously) $m=0$ at $%
h=h^{\ast }.$

\bibitem{projective} M. Kolesik, M.A. Novotny, P.A. Rikvold, and D.M.
Townsley, 
in \textit{Computer Simulation Studies in Condensed Matter Physics X}, D.P.
Landau, K.K. Mon, and H.B. Sch\"{u}ttler Eds., Springer Verlag, Heidelberg
(1997), pp.246-251.%

\bibitem{referee} We thank one of our referees for pointing out this subtlety. 

\bibitem{DSP} P.A. Rikvold and B.M. Gorman, in {\it Annual Reviews of Computational Physics},
edited by D. Stauffer, World Scientific, Singapore (1994), pp. 149-191.

\bibitem{Metropolis} A better definition of the effective temperature, with no 
singular behavior at $\Delta{\cal H} = 0$, may be obtained 
using the Metropolis rule, instead of Glauber rule, in the definition of rate 
(\ref{rate}).\cite{Pablotesis}


\end{thebibliography}
\end{document}